\documentclass[prb,aps,twocolumn,showpacs,floatfix,a4paper]{revtex4-1}
\usepackage[utf8x]{inputenc}
\usepackage[T1]{fontenc}

\usepackage{amsbsy}
\usepackage{geometry}
\usepackage{color}
\usepackage{graphicx}
\usepackage{color}
\usepackage{amsmath}
\usepackage{amssymb}
\usepackage{amssymb}
\usepackage{amsfonts}
\usepackage[caption=false]{subfig}
\usepackage{threeparttable}
\definecolor{darkgreen}{RGB}{0,160,48}

\newcommand{\etal}{\textit{et al.}}

\newcommand{\beq}{\begin{equation}}
\newcommand{\eeq}{\end{equation}}
\newcommand{\beqn}{\begin{eqnarray}}
\newcommand{\eeqn}{\end{eqnarray}}
\def\qq{\mathbf{q}}

\def\bise{Bi$_2$Se$_3$}
\def\EDT#1{#1} 

\begin{document}
\title{Thermal conductivity of Bi$_2$Se$_3$ from bulk to thin films: theory and experiment.}

\author{Lorenzo Paulatto}
\email{lorenzo.paulatto@sorbonne-universite.fr}
\affiliation{Sorbonne Universit\'{e}, CNRS, Institut de Min\'eralogie, de Physique des Mat\'eriaux et de 
Cosmochimie, IMPMC, UMR~7590, 4 place Jussieu, 75252 Paris Cedex 05, France}

\author{Danièle Fournier}
\author{Massimiliano Marangolo}
\author{Mahmoud~Eddrief}
\author{Paola~Atkinson}
\author{Matteo~Calandra}
\affiliation{Sorbonne Université, CNRS, Institut des NanoSciences de Paris, INSP, UMR 7588, F-75005 Paris, France}

\begin{abstract}
We calculate the lattice-driven in-plane $(\kappa_{\parallel})$ and out-of-plane $(\kappa_{\perp})$ thermal conductivities of Bi$_2$Se$_3$ bulk, and of films of different thicknesses, using the Boltzmann equation with phonon scattering times obtained from anharmonic third order density functional perturbation theory. 
\EDT{ We compare our results for the lattice component of the thermal conductivity with published data for $\kappa_{\parallel}$ on bulk samples and with our room-temperature thermoreflectance measurements of $\kappa_{\perp}$ on films of thickness (L) ranging from 18~nm to 191~nm, where the lattice component has been extracted via the Wiedemann-Franz law.
Ab-initio theoretical calculations on bulk samples, including an effective model to account for finite sample thickness and defect scattering, compare favorably both for the bulk case (from literature) and thin films (new measurements).}
In the low-T limit the theoretical in-plane lattice thermal conductivity of bulk Bi$_2$Se$_3$ agrees with previous measurements by assuming the occurrence of intercalated Bi$_2$ layer defects.
The measured thermal conductivity monotonically decreases by reducing $L$, its value is $\kappa_{\perp}\approx 0.39\pm 0.08$~W/m$\cdot$K for $L=18$ nm and $\kappa_{\perp}=0.68\pm0.14$~W/m$\cdot$K for $L=191$ nm. 
We show that the decrease of room-temperature $\kappa_{\perp}$ in Bi$_2$Se$_3$ thin films as a function of sample thickness can be explained by the incoherent scattering of out-of-plane momentum phonons with the film surface. Our work outlines the crucial role of sample thinning in reducing the out-of-plane thermal conductivity.
\end{abstract}

\maketitle

\section{Introduction}\label{introduction}

While the thermoelectric properties of Bi$_2$Te$_3$ have been widely studied both for bulk and
thin films\cite{Behnia2015}, interest in the isostructural topological insulator 
\bise{} mostly focused on its peculiar electronic structure and little is known on its thermal \EDT{conductivity}. The main reason is that its Seebeck coefficient is lower than that of Bi$_2$Te$_3$ and its thermal conductivity is somewhat higher, leading to a worse thermoelectric figure of merit $ZT$. Notwithstanding that, the situation could be different in Bi$_2$Se$_3$ thin films, where thermal conductivity could be reduced due to scattering with sample borders.\cite{Dresselhaus} Little is known of the thermal conductivity in this case.

\bise{}, similarly to  Bi$_2$Te$_3$, has a lamellar structure, consisting of sheets of covalently bonded Se-Bi-Se-Bi-Se atoms that are held together by weak interlayer van-der-Waals bonds. This highly anisotropic crystal structure is reflected in anisotropic thermal and electrical conductivity.  Thermal conductivity of bulk Bi$_2$Se$_3$ has been measured by several authors.
Navratil and coworkers \cite{navratil,navratil2} measured the thermoelectric properties and the
in-plane thermal conductivity in bulk Bi$_2$Se$_3$ and extracted the lattice contribution to $\kappa_{\parallel}$ (lying in the plane of the \bise{} layers, perpendicular to the $[111]$ direction). They find values of the order of $1.33-1.63$~W/m$\cdot$K at room temperature. Furthermore, the authors found a drop of $\kappa_{\parallel}$ at low temperatures which they attribute to the presence of charged Se vacancies. 

There is a certain variance in experimental values of $\kappa$: in Ref. \onlinecite{HorPhysRevB.79.195208} and for p-doped samples, the in-plane conductivity was found to be of the order of 
$1.25$ (W/m$\cdot$K), however the lattice contribution was not extracted. The room-temperature total lattice and electronic conductivity were also estimated in Ref.~\onlinecite{Tcond25} and found to be $2.83$~W/m$\cdot$K and $1.48$~W/m$\cdot$K, respectively. More recently, the total in-plane thermal conductivity was estimated to be $3.5$~W/m$\cdot$K.\cite{Fournier_2018}

In this work we present a detailed theoretical and experimental investigation of the thickness dependence of the out-of-plane thermal conductivity $\kappa_\perp$ for thin films grown by molecular beam epitaxy and having a thickness $L$ between 18 and 191~nm. We find that the out-of-plane thermal conductivity decreases monotonically with thickness. By using first-principles electronic structure calculations, we show that this is mostly due to the suppression of long-wavelength phonon propagating along the c-axis. Moreover, we show that the low temperature behaviour of the lattice component  of the thermal conductivity is mostly due to intercalated Bi$_2$ layers and not to Se vacancies, as suggested in previous works.\cite{navratil,navratil2}

In Sec.~\ref{exps} we present the experimental setup and techniques used to grow and measure the conductivity in thin films. In Sec.~\ref{theory} we review and extend the theory of phonon-driven thermal transport in finite crystals.  We report in Sec.~\ref{cmethod} the computational details, in Sec.~\ref{geom} the crystal geometry, and in Sec.~\ref{electk} we compute the electronic contribution to thermal conductivity via the Franz-Wiedemann law. In Sec.~\ref{results} we present our results on electronic structure (Sec.~\ref{pw}), phonon dispersion (Sec.~\ref{ph}), bulk thermal conductivity (Sec.~\ref{tk}) and thin films (Sec.~\ref{sec:tkthin}).

\section{Experiment}\label{exps}

\subsection{Thermal conductivity measurements}\label{method}
Thermal properties of thin films may be obtained at room temperature using modulated thermoreflectance microscopy\cite{rosencwaig,max-2}. \EDT{In this setup the heat diffusion associated with a heat source created by an intensity modulated pump beam is measured at the sample surface using the variation in the coefficient of optical reflection, which is measured by a probe beam impinging the heated area by a probe beam impinging the heated are}. The pump beam is a 532~nm Cobolt laser focused on the sample through a $\times$50 (0.5~NA) objective microscope. The pump laser is intensity-modulated in a frequency range 100~Hz $-$ 1 M~Hz. Since the light penetration depth is around 10~nm  a large amount of heat is released in the thin film. 

Surface temperature is then affected by heat diffusion carried by thermal waves\cite{rosencwaig,max-2}. The setup permits a spatial measurement of the surface temperature around the pump beam by a probe 488~nm Oxxius  laser that is reflected on the heated surface. The variations of the reflectivity (amplitude and phase) are directly proportional to the modulated temperature variation \EDT{through the refractive index variation and measured by a lock-in amplifier. To avoid artefacts due to variations in the optical quality of the surface, the probe beam is fixed on a small good quality area}, while the pump beam is scanned around the probe beam. Finally, the amplitude and phase experimental data are fitted according to a standard Fourier diffusion law to extract the thermal parameters (thermal conductivity $k$ and /or thermal diffusivity $D$):
\begin{align}
    D = k / d\,C\,,
\end{align}
where $d$ is the sample density and $C$ its specific heat\cite{max-3,max-4,max-5,max-6,max-7,max-8}.

\subsection{Thin films growth}

The growth of \bise{} thin films were conducted by Molecular Beam Epitaxy by following the procedure given in Ref.~\onlinecite{raman}. Flat $(111)$-B GaAs buffer surfaces were prepared in the III-V chamber and subsequently transferred to a second chamber where \bise{} thin films were growen with thicknesses ranging quen
om 18~nm up to 190~nm. The crystalline quality, epitaxial in-plane orientation, lattice parameter, crystalline structure of \bise{} films were verified by reflection high-energy electron diffraction and x-ray diffraction. \EDT{Surfaces are very flat ( $\sim 5$~nm rms roughness) and mirror-like facilitating the reflectivity measurements.}

\section{Theory}\label{theory}

\subsection{Scattering mechanisms in finite crystals}\label{sec:scat}
\begin{figure}
\includegraphics[width=0.5\textwidth]{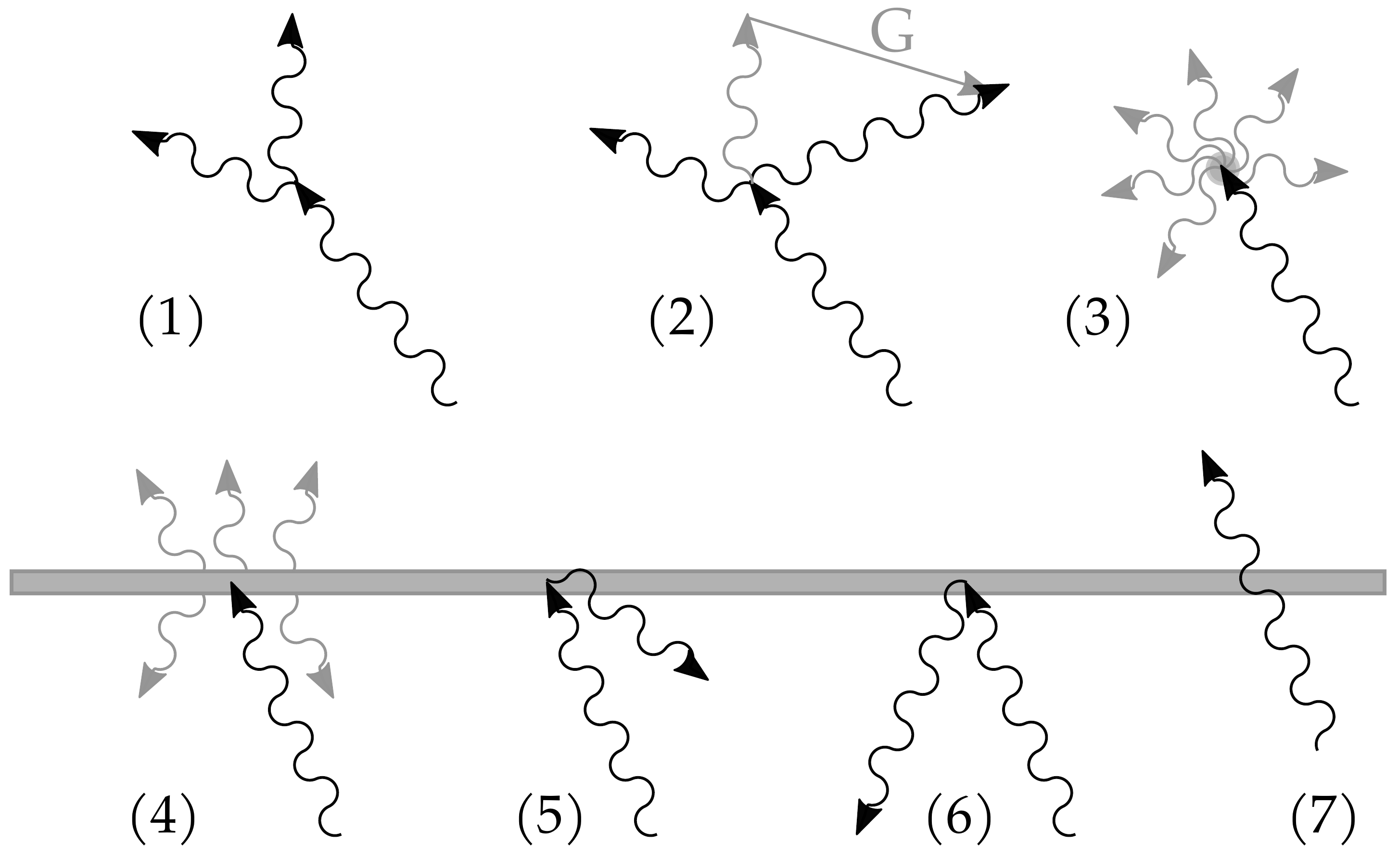}
  \caption{Possible scattering mechanisms in a slab-shaped crystal. 1) Normal momentum-conserving scattering (does not limit thermal transport). 2) \textit{Umklapp} scattering. Absorption by: 3) point defects or isotopic disorder, or 4) rough ``black'' surface, with black-body emission to maintain thermal equilibrium. 5) Reflection by a rough ``white'' surface. 6) Reflection by a smooth surface. 7) Transmission through an intercalated surface.}
  \label{fig:scatterings}
\end{figure}
The behaviour of lattice-driven thermal conductivity as a function of temperature has a typical shape which is mostly independent of the material : at high temperature it decreases as $1/T$, at lower temperature it has a maximum, then  going towards zero temperature it decreases sharply to a finite value. The behavior of the lattice thermal conductivity at high temperature is determined by the anharmonic phonon-phonon interactions\cite{peierls} with a contribution from defects. In the low temperature regime, named after Casimir who studied it in the 1930's\cite{casimir}, thermal conductivity is not a bulk property but it depends on the sample finite size.\cite{casimir}

Theoretical studies of the Casimir regime pre-date the possibility to study thermal conductivity by numerically integrating the phonon anharmonic properties\cite{casimir,berman53,berman55}. The standard approach consists in modeling the sample boundaries as black bodies that absorb a fraction of the colliding phonons, reflect the rest, and emit phonons to maintain thermal equilibrium. These works use geometric calculations, valid in the linear regime where only the acoustic phonons are taken into account, to predict the low-temperature thermal conductivity, usually with a single free parameter: the surface reflectivity. Invariably they assume a simple geometry for the crystal, such as a long cylinder or a long square parallelepiped with a temperature gradient between its opposite faces; in a shorter cylinder with polished faces the model requires the inclusion of multiple internal reflections\cite{berman55}.

With the arrival of more powerful numerical techniques, it has become more effective to model the surface at the phonon level, as a scattering probability. The probability of scattering from the boundaries can be combined with the probability of scattering due to phonon-phonon interaction in accordance with Matthiessen's rule.\cite{isotopes}. This approach has been used in more recent literature\cite{sparavigna} while keeping the assumptions of the long cylindrical geometry, not appropriate for application to thin films, where a temperature gradient can be applied orthogonally to the film lateral extension.

In Fig.~\ref{fig:scatterings} we have schematically depicted the possible scattering events responsible for limiting lattice-driven energy flow, we will briefly review them but for detailed discussion we redirect the reader to Ref.~\onlinecite{tk}. Mechanisms (1) and (2) are the intrinsic scattering processes: (1) is the ``normal'' (N) scattering, it conserves momentum and does not limit thermal conductivy. N scattering is only important at very low temperature (a few K). (2) The \textit{umklapp} (U) processes, that conserves crystal momentum modulus the addition of a reciprocal lattice vector, are the main limiting factor at high temperature and the prevalent intrinsic scattering mechanism. When studying thermal conductivity in the single-mode relaxation-time approximation (SMA), it is assumed that U scattering is dominant and that the scattered phonons are thermalized, i.e. that on average they are scattered toward the equilibrium thermal distribution. This approximation is very robust and works to within a few percent in a large range of temperatures and materials.\cite{nanolet}

Diagram number (3) depicts the Rayleigh scattering with a point defect, which could be a vacancy\cite{vacancies}, a substitutional defect or isotopic disorder\cite{isotopes}. Finally, events (4), (5) (6) and (7) are possible interactions between a phonon and the sample boundary: (4) is adsorption and re-emission by the surface; (5) is inelastic reflection by a ``white'' surface. We remark that, as it has been shown in Ref.~\onlinecite{berman55}, events (4) and (5) are equivalent from a thermal-transport point of view, we will just use (4) from here on. Further on, (6) is elastic reflection, where the momentum component parallel to the surface is conserved but the orthogonal component is inverted. Reflections can limit thermal conductivity in the direction orthogonal to the surface. Finally, in (7) a phonon can cross the boundary without scattering, this is of course not possible if the sample is suspended in vacuum, but can be the case if the sample is composed by multiple mis-matched segments, or if it contains stacking defects.

In order to describe the interface, we introduce three dimensionless parameters: the absorption fraction $f_a$, the reflection fraction $f_r$ and the transmission $f_t$, these are the probabilities that a phonon will undergo process (4), (6) or (7) respectively when it collides with the boundary. The condition $f_a + f_r + f_t = 1$ holds. In general these parameters may depend on phonon energy and its incidence angle, they can be computed using molecular dynamics techniques\cite{interfacial}. A special limit case is a very rough surface for which $f_a=1$.

\subsection{Thermal transport in the single-mode approximation}
In the single mode approximation (SMA) the thermal conductivity matrix is:
\begin{align}
 \kappa_{\alpha\beta} = \frac{\hbar^2}{N_0\Omega k_B T^2}\sum_j v_{\alpha,j} v_{\beta,j} \omega_j^2 n_j (n_j +1)\tau_j \label{eq:sma}
\end{align}
Where $j$ is a composite index running over the phonon wavevectors {\bf q} in reciprocal space and the phonon bands $\nu$; $N_0$ are the number of {\bf q}-points used to sample the Brillouin zone, $\Omega$ is the unit-cell volume, $k_B$ is the Boltzmann constant and $T$ is temperature. Inside the sum, the composite index $j$ stands for the band index $\nu$ and the wavevector $\qq$; then $\omega_j = \omega_\nu(\qq)$ is the phonon frequency, $v_j = \nabla_\qq \omega_\nu(\qq)$ is the phonon group velocity; $\alpha$ and $\beta$ are cartesian directions ($x, y, z$) $n_j = n(\omega_\nu(\qq))$ is the Bose-Einstein distribution and $\tau_j$ is the phonon relaxation time, or inverse full-width half-maximum\cite{lw}.

The SMA is accurate when umklapp or dissipative scattering is dominant over "normal" and elastic scattering, we have checked that this is always the case in \bise{} above 1~K: above this temperature the exact solution of the Boltzmann transport equation (BTE)\cite{tk} only increase $k$ by a couple percent. As the SMA equation is much cheaper to compute, easier to manipulate and has a more straightforward interpretation, we will use it exclusively in the rest of the paper.

\subsection{Thermal transport in thin film crystals}\label{sec:theofilms}
In order to progress further we have to take into account the real geometry of our sample. In this paper we will consider two cases: (i) a thin film of \bise{} of thickness $L$ along direction $z$ and virtually infinite in the other two directions with two very rough opposing surfaces; (ii) bulk \bise{} intercalated with partial planes of Bi$_2$, which is a common kind of crystal defect,\cite{raman, defects} at an average distance $L$.

In case (i) we consider a phonon emitted from a surface that moves toward the opposite surface with a $z$ component of its group velocity $v_z$. After a time $L/v_z$, the phonon will reach the other surface and be absorbed with probability $f_a$, giving the first phonon scattering rate $\gamma_a^{(1)}$ and the relaxation time $(\tau_a^{(0)})^{-1}=f_a \frac{v_z}{L}$.
If it is not absorbed (probability $f_r = 1-f_a$), the phonon will be reflected back toward the initial surface with identical speed and it will undergo a second absorption/reflection process. \EDT{The probability of a third reflection is $f_r^2$, for the $n$th reflection is is $f_r^{(n-1)}$. After summing the geometric series, the total effective lifetime is}:
\begin{align}
 \nonumber\\
 (\tau_a)^{-1} = 2\gamma_a  =& f_a \frac{v_z}{L} \sum_{i=1,n} f_r^{i-1} = \frac{v_z}{L}\left(\frac{f_a}{1-f_r}\right) \,\label{eq:taua}.
\end{align}
For boundary scattering, $f_a = 1-f_r$ conveniently cancels out giving $\tau_a = \frac{L}{v_z}$, but we prefer to leave eq.~\ref{eq:taua} in a general form to consider more general cases. Furthermore, if a phonon is reflected, its velocity component that is orthogonal to the surface will be inverted. We can account for this possibility in eq.~\ref{eq:sma} renormalizing $v_z$ in the following way: a fraction $f_r$ of the phonons will change the sign of $v_z$, a fraction of them will hit the opposite boundary, be reflected a second time and change sign again, and so on. The material is traversed in a ``flying'' time $\tau_f = L/v_z$. During this time phonons are scattered at a rate $P_x = \tau/\tau_f$, resetting the process. With $\tau$ being its total (intrinsic and extrinsic) relaxation time. This can be expressed as:
\begin{align}
  \tilde{v_z} =& \sum_{i=0,\infty} (-\frac{\tau_f}{\tau} f_r)^i v_z 
  = v_z \left( 1+ \frac{\tau}{\tau_f}f_r \right)^{-1}\,\label{eq:vz}.
\end{align}
Again, we do not replace $\tau_f$ with $1-\tau_a$ because we want to keep this equation as general as possible.

In case (ii), a bulk material intercalated with planes, the reasoning is very similar, with the caveat that $1-f_a=f_r+f_t$, although for an atom-thin intercalated layer we can safely assume that $f_r$ is almost zero.

The final formula for $\kappa$ becomes:
\begin{align}
 \kappa_{\alpha\beta} = \frac{\hbar^2}{N_0\Omega k_B T^2}\sum_j \tilde{v}_{\alpha,j} \tilde{v}_{\beta,j} \omega_j^2 n_j (n_j +1)\tau^{tot}_j \label{eq:kappa-final}
\end{align}
With $\tilde{v}$ from eq.~\ref{eq:vz}, and $\tau$ comprises all the scattering terms, summed with the Matthiessen's rule:
\begin{align}
\tau^{tot} = \left(  \tau_{ph-ph}^{-1} + \tau_a^{-1} +... \right)^{-1}
\label{eq:tautot}
\end{align}
where additional scattering terms like point-defect scattering, can be added. In the case of \bise{} we will see in Sec.~\ref{sec:tkthin} the impact that internally reflected phonon have on the conductivity. We also underline that while our samples are not free-standing in the experiment, we assume that, on the time-scale of the measurements, the transmission of heat from \bise{} to the GaAs substrate can be ignored.

\section{Technical details of first-principles simulations}\label{sec:technical}
\subsection{Computational method}\label{cmethod}
All calculations have been performed using the \textsc{Quantum-ESPRESSO} suite of codes\cite{qe1,qe2}, and in particular \textsc{ph}\cite{ph} and \textsc{d3q}\cite{lw} modules. \textsc{D3q} efficiently computes 3-body anharmonic force constants from density functional perturbation theory\cite{dfpt1,dfpt2,dfpt3} and the "2n+1" theorem\cite{dn1,dn2}. We also used the related \textsc{Thermal2} codes\cite{lw,tk} to compute intrinsic phonon lifetime, scattering with isotopic defects\cite{isotopes} and border effects and to compute lattice-driven thermal conductivity in the single-mode relaxation time approximation (RTA) and by solving iteratively the full Peierls-Boltzmann\cite{peierls} transport equation via a functional minimization\cite{tk}.

We used the generalized gradient approximation with the  PBE\cite{pbe} parametrization. We employed norm-conserving pseudopotentials from the SG15-ONCV library\cite{oncv1,oncv2,oncv3}, which include scalar-relativistic effects, and custom made pseudopotentials based on the SG15-ONCV pseudisation parameters, but including full-relativistic spin-orbit coupling (SOC). The \textsc{ph} code includes SOC effects\cite{phso}, but the \textsc{d3q} code does not, hence we always used scalar-relativistic pseudopotentials for the third order calculations.

We used a kinetic energy cutoff of 40~Ry for the plane wave basis set, and we integrated the electronic states of the Brillouin zone (BZ) using a regular Monkhorst-Pack grid of $8 \times 8 \times 8$ k-points, except when computing the effective charges and static dielectric constant via linear response, which only converged with a much finer grid of $32 \times 32 \times 32$ points.

The phonon-phonon interaction was integrated using a very fine grid of $31 \times 31 \times 31$ q-points, when computing the linewidth along high-symmetry direction or the phonon spectral weight. A coarser grid of $19 \times 19 \times 19$ points was used when computing the thermal conductivity. The SMA thermal Boltzmann equation was itself integrated over a grid $19 \times 19 \times 19$ q-points. All the grids in this paragraph were shifted by a random amount in order to improve convergence avoiding symmetry-equivalent points. The finite size effects of section \ref{sec:theofilms} where included in the calculation of $\kappa$ using an in-house Octave\cite{octave} code available upon request.

\subsection{Simulated crystal structure}\label{geom}

Bi$_2$Se$_3$ belongs to the tetradymite-type crystal with a rhombohedral structure (point
group R$\bar{3}$m, Wyckoff number 166). In the rhombohedral unit cell there are three Se and two Bi atoms. One Se atom is at the (1a) site $(0,0,0)$, the remaining Se and Bi atoms are at the two-fold (2c) sites of coordinates $(u,u,u)$ and $(-u,-u,-u)$, with one free parameter for each species, which we indicate as $u_\mathrm{Se}$ and $u_\mathrm{Bi}$ respectively.
Structural parameters obtained from experimental measurement and from \textit{ab-initio} simulations are shown in Tab. \ref{tab:lattice}.
Both the scalar relativistic and the fully relativistic PBE calculations give substantially expanded structures, both in $a$ and $c$, as is customary in this approximation. As the electronic band gap is quite sensitive to the geometry, we also calculated the bandgap for the experimental volume. In all fully relativistic calculations the gap is substantially overestimated, in agreement with previous theoretical calculations.

\begin{table*}
\begin{threeparttable}
    \caption{Structural parameters of bulk \bise{} : lattice parameters, $a$, $c$, unit cell volume $V$, lattice positions of Se and Bi, $u_{Se}$, $u_{Bi}$, and the bulk band gap. Experimental data (Ref.~\onlinecite{expgap}) is given, and compared with the scalar relativistic (SR) and fully relativistic (SOC) calculations of this work where theoretically determined or experimentally determined lattice parameters were used as model input. Last row: results from literature DFT simulation. 
    }
\begin{tabular}{|l| l| cc c| cc| c|}
\hline
 & structure ~~~~ & ~a (\r{A}) ~ & ~c (\r{A}) ~ & ~V (\r{A}$^3$)~ &
   ~~ $u_\mathrm{Se}$~~ & ~~$u_\mathrm{Bi}$~~ & ~gap (meV)~ \\
\hline
 \multicolumn{2}{|c|}{ Experimental:}  
    & 4.138 & 28.64 & 422.8 & - & - & 200$\pm$5 \\
\hline
\multicolumn{8}{c}{}\\
\hline
approximation & structure & ~a (\r{A}) ~ & ~c (\r{A}) ~ & ~V (\r{A}$^3$)~ &
   ~~ $u_\mathrm{Se}$~~ & ~~$u_\mathrm{Bi}$~~ & ~gap (meV)~ \\
\hline
\multicolumn{8}{c}{this paper:}\\
\hline
SR       & theoretical & 4.198 & 30.12 & 459.6 & 0.217 & 0.398 & 471 \\
SOC      & theoretical & 4.211 & 29.76 & 457.0 & 0.215 & 0.399 & 373 \\
SR       & experimental\tnote{a} & 4.138 & 28.64 & 422.8 & 0.211 & 0.400 & 183 \\
SOC      & experimental\tnote{a} & 4.138 & 28.64 & 422.8 & 0.209 & 0.401 & 333/444\tnote{b} \\
\hline
\multicolumn{8}{c}{DFT simulations in literature:}\\
\hline
SOC/\textsc{vasp}   & experimental\cite{bise2012-prl}   & 4.138  & 28.64  & 422.8 & - & - & 320 \\
\hline
\end{tabular}
\begin{tablenotes}
\item[a] Lattice parameters from Ref.~\onlinecite{wyck} as in \onlinecite{bise2012-prl}, also compatible with Ref.~\onlinecite{latparoft}.
\item[b] Indirect/direct gap.
\end{tablenotes}
    \label{tab:lattice}
\end{threeparttable}
\end{table*}

Bi$_2$Se$_3$ thin films measured here had thicknesses between 18~nm and 191~nm. The stack of 2 Bi and 3 Se atoms (one Q-layer) is $0.984$~nm thick, which means that even for the thinnest slab we have 18 or 19 Q-layers.

\subsection{Electronic contribution to thermal conductivity}\label{electk}
Commercially available bulk \bise{} samples, and our epitaxial \bise{} thin film samples, are always doped to a certain extent by the presence of vacancies and imperfect stoichiometry. For this reason, even if the perfect material is a small gap insulator, we expect to measure a certain amount of electron/hole driven thermal and electric transport. Although there is no simple way to isolate the electronic contribution to thermal transport ($k_e$), we can estimate it from electric conductivity.

Mobility ($\mu$), doping and electrical resistivity ($\rho_\parallel$) of most of the thin films have been measured by conventional Hall bar measurements performed in the plane perpendicular to the trigonal axis, $c$, of \bise{} thin films. 

It turns out that, at 300~K our thin films are $n$-doped and present a metallic behavior with a carrier concentration between $1-2\cdot 10^{19}$~cm$^{-3}$,
$\mu \sim 300-400$~cm$^2/$V$\cdot$s and $\rho_{\parallel} \sim 1-1.5$~m$\Omega\cdot$cm.
A coarse evaluation of the electronic contribution to the in-plane thermal conductivity, $k_{e,\parallel}$, 
can be given by using the Wiedemann-Franz law $k_{e,\parallel} =LT/\rho_{\parallel}$, 
with $L$ ranging between
$2$ and $2.2\cdot 10^{-8}$ V$^2$K$^{-2}$.\cite{navratil}
Consequently, $k_{e,\parallel}$ at 300~K ranges between $0.4-0.7$~W/m$\Omega\cdot$K. 

Concerning the out-of-plane resistivity term, $\rho_{\perp}$, a rough estimation can be given by $\rho_{\perp} \sim 3.5 \rho_{\parallel}$ since such a ratio is found in bulk \bise{}\cite{stordeur} leading to $k_{e,\perp}$ between $0.1-0.2$~W/m$\cdot$K. 
It is worthwhile to underline that $\rho_{\parallel}$ measurements performed in very thin \bise{} samples (9 QL) give a slight lowering of the resistivity (0.7 m$^2\cdot$cm),  which may be caused by $k_{e,\perp}$ slightly increasing at very low thickness.

\section{Results and discussion}\label{results}

\subsection{Electronic structure}\label{pw}
\begin{figure*}
  Theoretical lattice parameter:
  \subfloat[a][SR - 471~meV]{\includegraphics[width=0.32\textwidth]{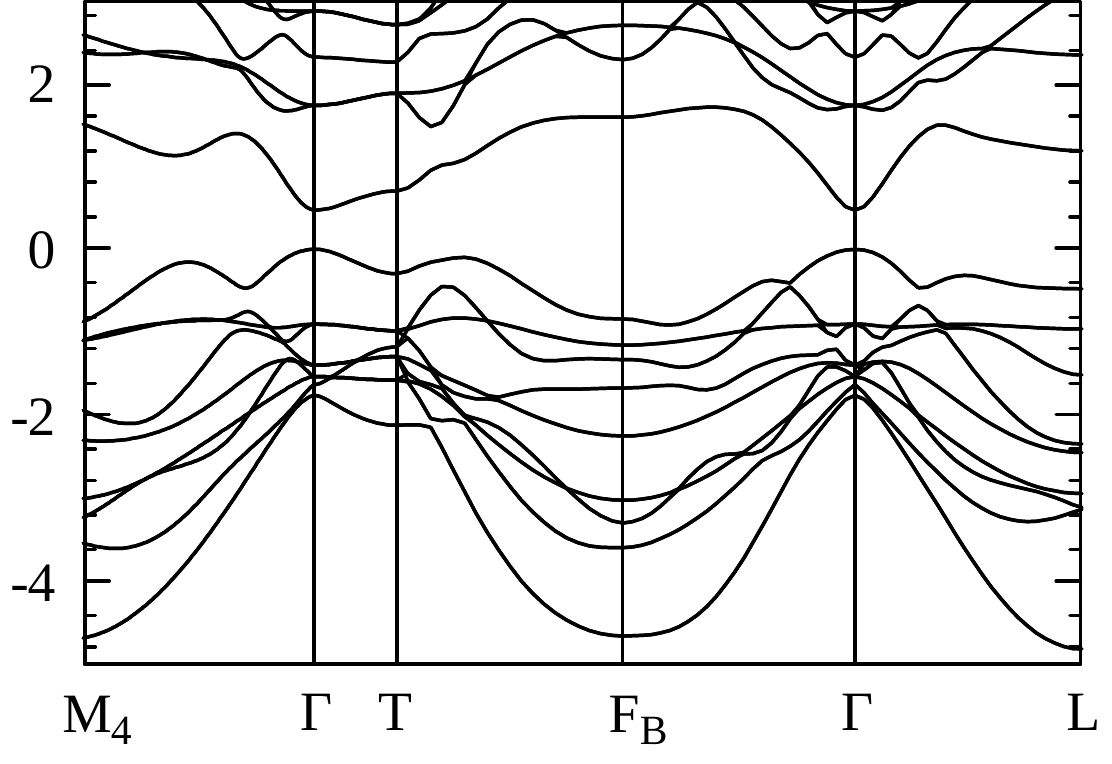}}
  \subfloat[b][SOC - 373~meV]{\includegraphics[width=.32\textwidth]{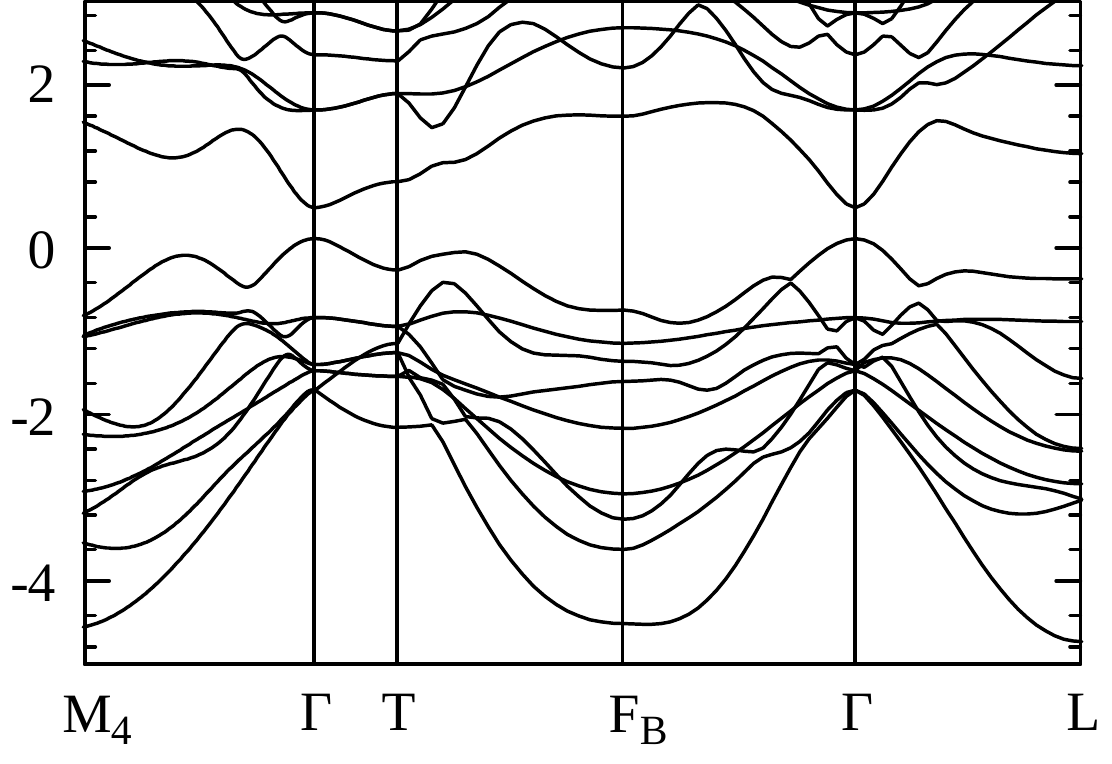}}
  \subfloat[c][Gap detail ]{\includegraphics[width=0.32\textwidth]{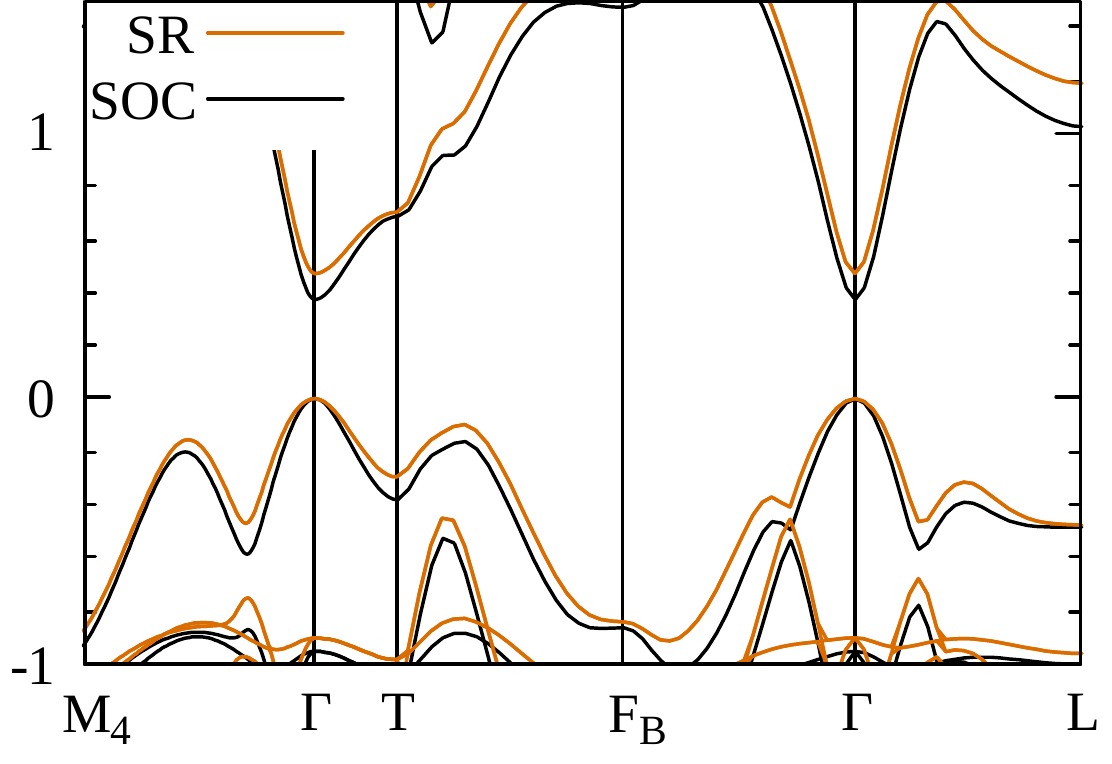}}
  \vspace{.5cm}
  Experimental lattice parameter:
  \subfloat[d][SR - 183~meV]{\includegraphics[width=0.32\textwidth]{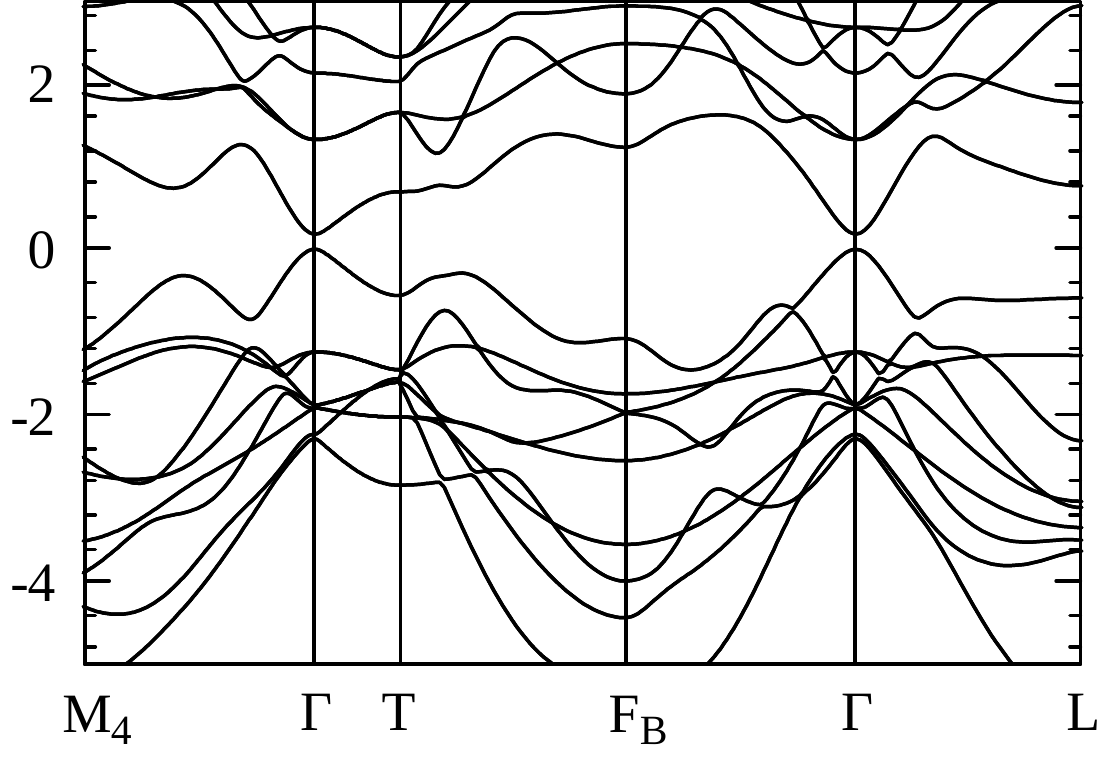}}
  \subfloat[e][SOC - 333~meV]{\includegraphics[width=.32\textwidth]{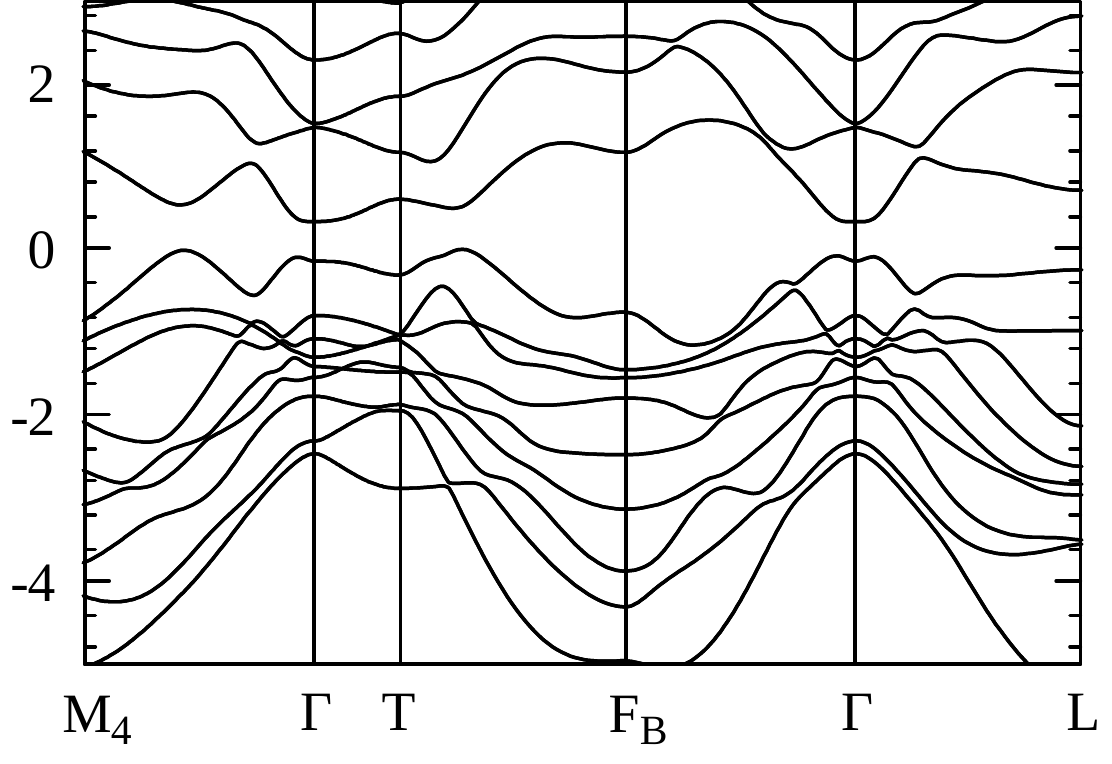}}
  \subfloat[f][Gap detail ]{\includegraphics[width=0.32\textwidth]{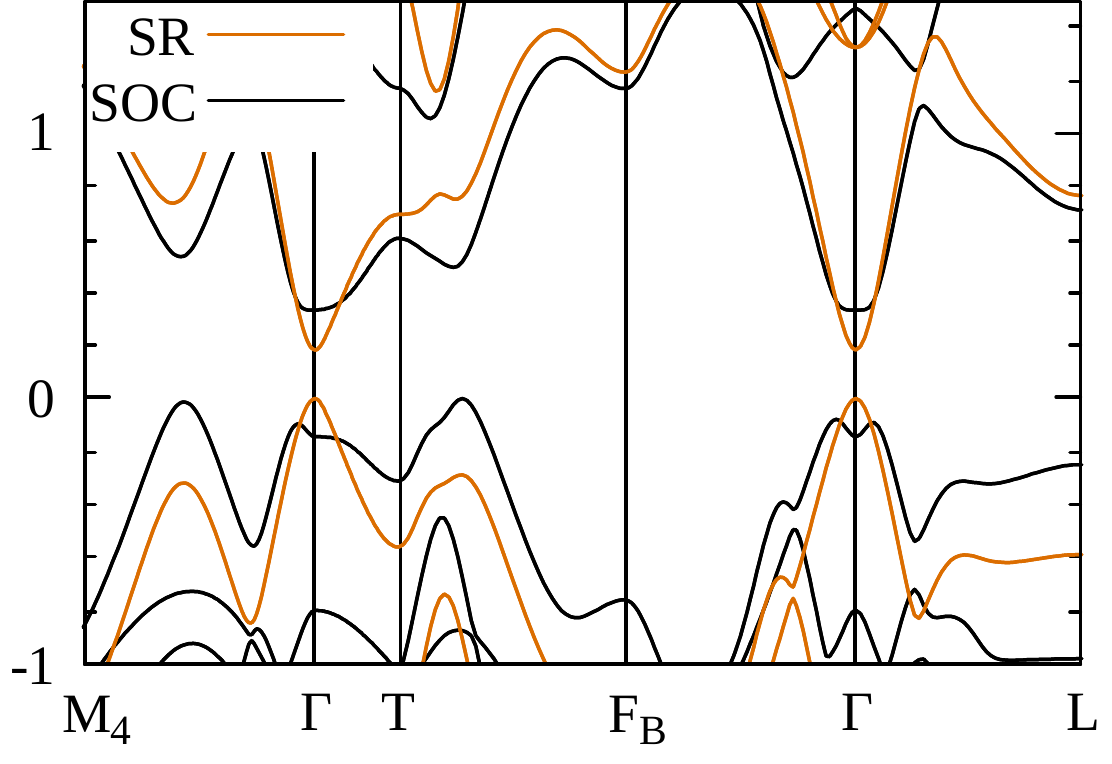}}

  \caption{Calculated electronic band structure of bulk Bi$_2$Se$_3$. In the first row: using each method's respective calculated theoretical lattice parameter; second row: experimental lattice parameter. First column: scalar relativistic (SR),  second column: fully relativistic (SOC), third column: comparison SR/SOC close to the Fermi energy.}
  \label{fig:elbands}
\end{figure*}
Before delving in the calculation of the vibrational properties, we have taken care to verify that the approximations used to deal with the \textit{ab-initio} problem are valid. A few potential difficulties have to be accounted for: the first is that \bise{} is a small gap semiconductor, special care is needed  in order to prevent the gap from closing. The gap magnitude depends both on the kind of local- or semi-local- exchange and correlation kernels and on the lattice geometry used. In any case, the Kohn-Sham single particle gap is not guaranteed to be correct, as it is not a ground-state property. However, if its character is different from the experimental one it can indicate that the employed approximation is non appropriate. The optical gap has been measured experimentally as being direct and $200 \pm 5$~meV wide\cite{expgap}. 

Another difficulty is treating the inter-layer Van der Waals bond; the inter-layer distances are usually over-estimated by standard local density approaches. However it is possible to improve the agreement with experiment of vibrational frequencies using the experimental lattice parameters. For this reason, we have simulated the electronic structure using both the experimentally measured lattice parameter and the theoretically calculated values. In both cases we optimize the internal degrees of freedom to avoid unstable phonons.

In Fig.~\ref{fig:elbands} we plot the electronic band structure, in a range of a few eV around the Fermi energy. We note that the band structure is very similar except for the case where relativistic effects are included in conjunction with the experimental lattice parameter (Figure \ref{fig:elbands}b). When using the theoretical lattice parameter, (values give in Table \ref{tab:lattice}) the gap is of 471~meV in the scalar relativistic case and it decreases to 373~meV in the fully relativistic calculation. Conversely, if we use the experimental volume and optimize the internal coordinates, the gap is smaller for the scalar relativistic calculation 183~meV (Fig.~\ref{fig:elbands}c) while in the fully relativistic case the indirect gap is at 333~meV and the direct gap at 444~meV (Fig.~\ref{fig:elbands}d).

\subsection{Phonon dispersion}\label{ph}
\begin{figure}
\includegraphics[width=0.5\textwidth]{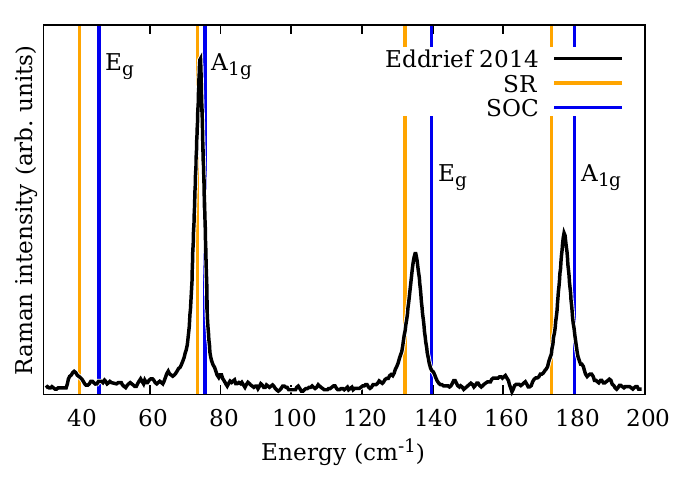}
  \caption{Experimental\cite{raman} Raman spectrum compared with theoretical calculations. The vertical lines indicate the theoretical harmonic phonon frequencies SR (orange) or with SOC (blue) at the experimental lattice parameter.}
  \label{fig:raman}
\end{figure}

\begin{figure}
\includegraphics[width=0.5\textwidth]{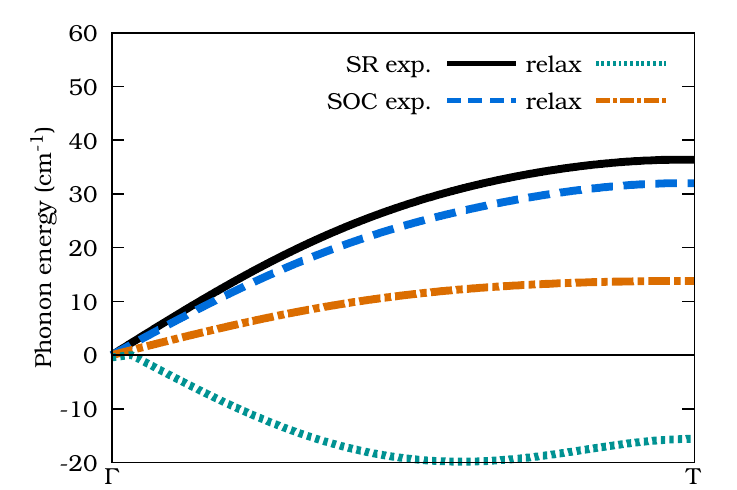}
  \caption{Detail of the longitudinal-acoustic phonon dispersion along the [111] direction (parallel to the $c$ axis), showing how the effect of the lattice parameter. We plot four cases: at experimental lattice parameter without SOC (black solid) and with SOC (blue dash); at the relaxed theoretical lattice parameter: without SOC (green dot) and with SOC (orange dash-dot)}
  \label{fig:ph-detail}
\end{figure}

The phonon dispersion does not change dramatically with the different choices of SOC treatment, however we do observe a global, relatively constant, rescaling of the frequencies which can be associated with the difference in the unit cell volume. When not including SOC, if the theoretical lattice parameter is used, the phonon dispersion exhibits negative frequencies. However, when using the experimental lattice parameters, which correspond to a 5\% smaller unit cell volume, this instability is removed. Furthermore, if SOC is included, both theoretical and experimental lattice parameters yield stable phonons, as long as the internal degrees of freedom are properly relaxed. 
Because inter-planes binding is mediated by Van der Waals forces, which usually leads to an over-estimation of the bond length in the PBE approximation, we expect that the modes changing the inter-layer distance will be too soft. 
For this reason, we show in Fig.~\ref{fig:ph-detail} the phonon band that is more sensitive to a change in lattice parameter, it is the longitudinal acoustic (LA) mode along the [111] direction which, in the trigonal geometry, is orthogonal to the planes of Bi or Se. 

We have compared the phonon frequencies at the $\Gamma$ point with available data from infrared and Raman spectroscopy, in order to establish which method gives a closer match to the phonon frequencies. These comparisons are summarized in table~\ref{tab:phonons}. We note that using the experimental lattice parameter gives a consistently better match for the Raman active modes than using the theoretical one, hence we will focus on the former. The SR calculations slightly under-estimate the phonon frequencies, while with the inclusion of SOC the theoretical frequencies tend to over-estimate the measured ones. SOC is more accurate for the highest optical bands, but less so for the low-energy bands; as the latter are more important for thermal transport, due to the Bose-Einstein factor of eq.~\ref{eq:sma}, we expect that not including SOC may give a better match for the value of $\kappa$. We verified that our calculations at the theoretical lattice parameter agree with those of Ref.~\onlinecite{cheng11}.

\subsubsection{Effective charges and LO-TO splitting}\label{loto}

\begin{figure}
\includegraphics[width=0.5\textwidth]{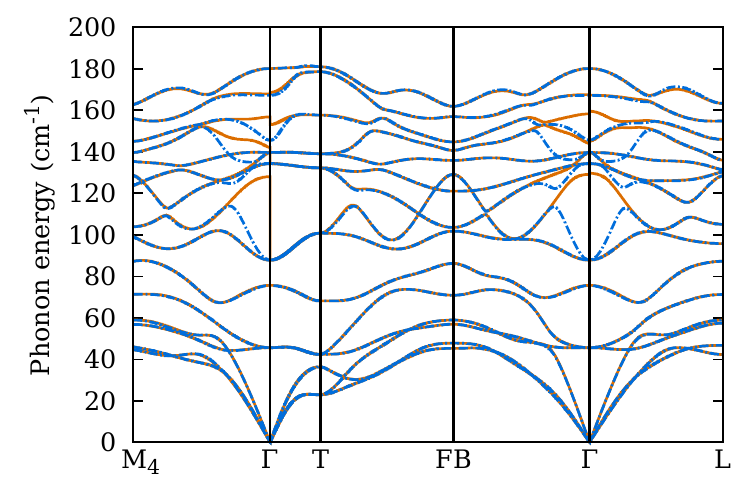}
  \caption{Phonon dispersion with (orange fullline) and without (blue dash-dot line) long-range LO-TO splitting mediated by effective charges. These simulations use the experimental lattice parameters, PBE, SR.}
  \label{fig:loto}
\end{figure}

Even high-quality \bise{} single crystals are doped by a relatively large amount ($10^{18}-10^{19}$~e$/$cm$^3$) of lattice defects. Below we will see that this doping has no significant effect on the electronic structure nor on the geometry of the crystal, but it is sufficient to effectively screen long-range/small-wavevector splitting of longitudinal and transverse optical modes (LO-TO splitting). In figure \ref{fig:loto} we have plotted the phonon dispersion with and without the LO-TO splitting. We can see that a couple of modes are particularly affected. If we number the modes in order of increasing energy, these are the modes 8 and 9 where the atoms move in the plane perpendicular to [111] with Se ions going in the direction opposite to that of Bi ions. 
These modes are degenerate without LO-TO at an energy of 87~cm$^{-1}$, but the mode which is aligned with the {\bf q}-vector jumps to 130~cm$^{-1}$ when long-range effects are included. When coming from the [111] direction itself, along the $\Gamma$-$T$ line, the two modes remain degenerate, because the {\bf q} wavevector is parallel to the polarisation.

\begin{table*}
\begin{threeparttable}
    \caption{Phonons frequency at $\Gamma$, comparison of experimental data and calculations. }
\begin{tabular}{|l| cccc| cccc|}
\multicolumn{9}{c}{experiments:}\\
\hline
~ & ~ & \multicolumn{3}{c|}{IR active (E$\parallel$c)} & \multicolumn{4}{c|}{Raman active} \\
Symmetry & ~~$E_u$~~ & ~~$E_u$~~ & ~~$A_{2u}$~~ & ~~$A_{2u}$~~ 
         & ~~$E_g$~~ & ~~$A_{1g}$~~ & ~~$E_g$~~ & ~~$A_{1g}$~~ \\
\hline
Ref. \onlinecite{infrared} (50~K)\tnote{a}~  &
61 & 134 & $-$ & $-$ & $-$ & $-$ & $-$ & $-$ \\
Ref. \onlinecite{infrared} (300~K)\tnote{a}~  &
65 & 129 & $-$ & $-$ & $-$ & 72 & 131.5 & 174.5 \\
Ref. \onlinecite{raman}   &
$-$ & $-$ & $-$ & $-$ & 39 & 74 & 135 & 177 \\
\hline
\multicolumn{9}{c}{simulations harmonic level:}\\
\hline
SR    theo.         & 76.2-123.8 & 129.7 & 143.4 & 161.2 & 38.3 & 61.8 & 130.5 & 171.5 \\
SOC   theo.         & 64.7-111.2 & 123.6 & 135.6 & 154.7 & 38.7 & 63.3 & 121.6 & 166.3 \\
SR    exp.\tnote{b} & 87.8-129.8 & 134.2 & 145.2 & 166.4 & 45.6 & 75.6 & 139.7 & 180.1 \\
SOC   exp.\tnote{b} & 78.0-123.0 & 128.5 & 138.0 & 162.3 & 40.2 & 73.6 & 132.2 & 173.6 \\
\hline
\end{tabular}
\begin{tablenotes}
\item[a] Infrared peak position are at 50~K and 300~K respectively.
\item[b] Lattice parameters from Ref.~\onlinecite{wyck}
\end{tablenotes}
    \label{tab:phonons}
\end{threeparttable}
\end{table*}

\begin{figure}
\includegraphics[width=0.5\textwidth]{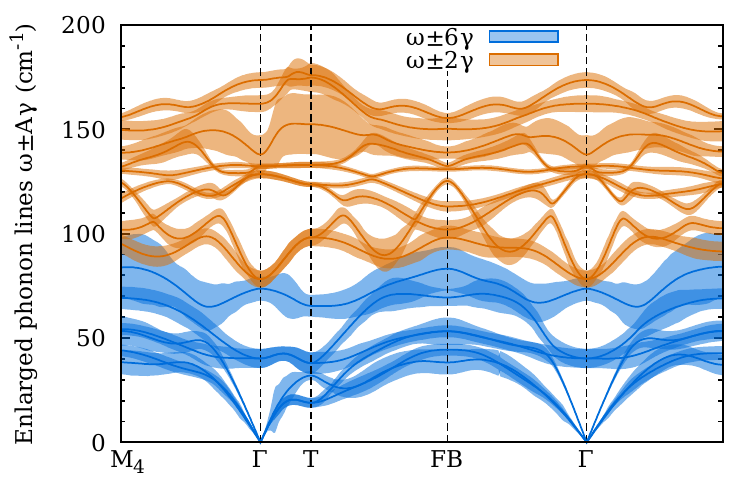}
  \caption{Phonon dispersion, the width of the lines is proportional to their intrinsic linewidth, a different prefactor was used before and after the phonon gap to improve readability: 2 for the first 6 bands, 6 for the following 9 bands.}
  \label{fig:disp}
\end{figure}

\subsection{Thermal transport in bulk}\label{tk}
\begin{figure*}
\includegraphics[width=0.5\textwidth]{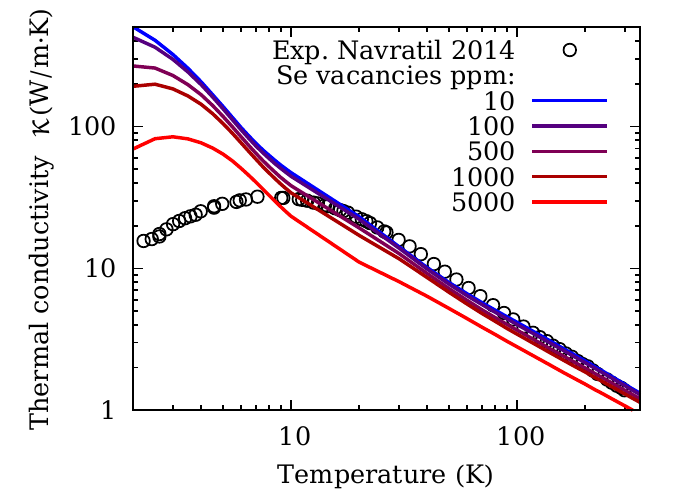}\includegraphics[width=0.5\textwidth]{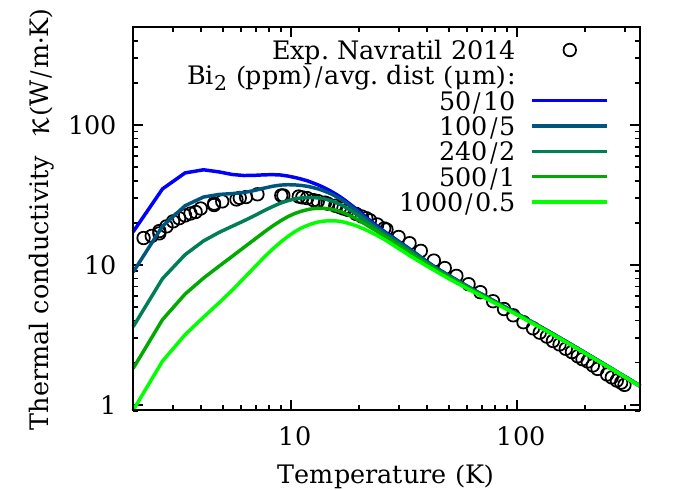}
  \caption{In plane (i.e. orthogonal to $c$ axis) $\kappa$ measured by Navratil\cite{navratil} (rounds) compared to simulations including the effect of Se vacancies (left) and Bi$_2$ layers intercalations (right) defects on $\kappa$, excess Bi$_2$ is expressed as average inter-plane distance and as ppm. \EDT{The behavior at very low temperature (i.e. below $2$~K) is not shown as it could suffer from the finite k-point mesh used in the calculations, due to the difficulty of converging the conductivity in the zero temperature limit}}.
  \label{fig:tk-defects}
\end{figure*}

\begin{figure}
\includegraphics[width=0.5\textwidth]{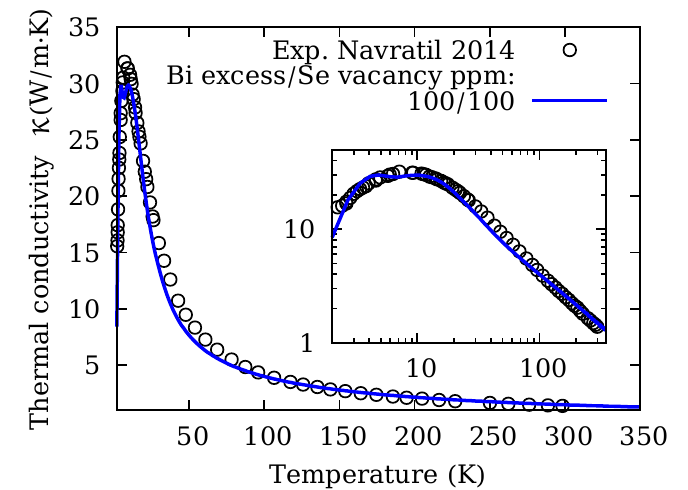}
  \caption{description Our b of the thermal conductivity with Ref.~\onlinecite{navratil} is obtained assuming around 100 ppm of Bi$_2$ partial layers and 100 ppm of Selenium vacancies.}
  \label{fig:tk-bulk-defects}
\end{figure}Ref.~Wonlinee have initially studied the phonon-driven thermal conductivity in the bulk phase as experimental data are available with good precision over a wide range of temperature. In particular we have taken as reference the data of Navratil and coworkers\cite{navratil}, where they estimate the fraction of lattice-driven and electron-driven transport. 

In Fig.~\ref{fig:tk-defects} we plot the experimental data of the in-plane  thermal conductivity $\kappa_{\parallel}$ measured in Ref. \onlinecite{navratil}, side by side with calculations from 2~K up to 400~K, in the RTA (we checked that the exact inversion of the Boltzmann transport equation yield practically identical values). As it can be seen, the room temperature behaviour of the lattice contribution to $\kappa_{\parallel}$ is in perfect agreement with our calculation of the intrinsic thermal conductivity. This agreement is possible thanks to the inclusion of lattice defects in the model, as explained in the rest of this section.

Below 20~K, $\kappa_{\parallel}$ is limited by extrinsic scattering processes such
as the scattering with sample borders, with the isotopes or/and with lattice defects. As Navratil \etal{} used a large mono-crystal, and isotopical effects are negligibles in \bise{},\footnote{The relative mass variance of Selenium isotopes is only $0.046\%$, while Bismuth has only one stable isotope} only lattice defects can explain the low temperature behaviour.

According to literature,\cite{defects} two kind of defects are common in \bise{} crystals: point-defect vacancies of Se, and Bi$_2$ partial-layer intercalation. Each Selenium vacancy contributes around two charges to the total doping, which means that at a doping concentration of around $10^{18}-10^{19}$~e$/$cm$^3$ the fraction of missing Se atoms is of order $100-1000$~ppm. We have simulated this defect concentration using an effective Rayleigh point-scattering model, \EDT{assigning to each vacancy an effective mass as in Ref.~\onlinecite{vacancies}}. We found that it is far too low to explain the low-temperature drop in thermal conductivity. Even taking an unrealistically high point-defect concentration, such as $50\,000$~ppm (5\%), the correct curve shape at low temperature is not reproduced. \EDT{Finally, we remark that using a more accurate, i.e. \emph{ab-initio}, estimate of the defect cross-section would be equivalent to a change in defect concentration, but would not change the shape of the curve.}

On the other hand, if we assume the presence of Bi$_2$ partial layers, we can include it in the simulation using Sparavigna-Casimir scattering theory, i.e. \EDT{using an effective model that} includes a scattering time which is proportional to the ratio between the phonon mean free path and the sample size. We tuned the average inter-defect distance to fit the temperature of maximum $\kappa$, around 10~K. The theoretical position of the maximum is a better fitting parameter than its absolute value, as the latter is very difficult to converge at low temperature in simulations. Notwithstanding that, the calculation reproduces the absolute value quite well, which strengthens the validity of our assumption. In Fig~\ref{fig:tk-bulk-defects}, the best agreement is found when the average distance between Bi$_2$ planes is fixed at 5~$\mu$m. Comparing this value to the size of the unit cell along $c$ gives a concentration of excess Bismuth of around $100$~ppm, and considering that each additional Bi atom provides three charges, this is compatible with the measured doping concentration.

We note that the effect of Selenium vacancies and Bismuth partial layer intercalation is qualitatively different: a increasing concentration of Selenium vacancies causes a global reduction of $\kappa$, on the other hand increasing the frequency of Bismuth partial layer intercalation moves the maximum of $\kappa$ toward higher temperatures, without changing its high-T value. If we combine the two types of defects, we observe that Selenium vacancies have virtually no effect until their concentration is greater than 100ppm, after which scattering from vacancies lead to a considerable reduction in $\kappa$ at higher T. As a consequence, the best match remains a concentration of around 100~ppm Bismuth partial layer intercalation with 100~ppm or less Selenium vacancy. This is compatible with the high n-type concentration ($\simeq10^{19}$~cm$^{-3}$) of the bulk material measured.

\subsection{Thermal transport in thin films}\label{sec:tkthin}

\begin{table}
\begin{tabular}{|c|c|c|}
    \hline
            ~ & \multicolumn{2}{c|}{Thermal conductivity:} \\
    Thickness &  Total (measured)  & Lattice (estimated)  \\
     (nm)     & {(W/m$\cdot$K)} & {(W/m$\cdot$K)} \\
     \hline
18 & 0.39 & 0.19 \\
30 & 0.52 & 0.32 \\
53 & 0.53 & 0.33 \\
105 & 0.56 & 0.36 \\
191 & 0.68 & 0.48 \\
\hline
\end{tabular}
\caption{Out-of-plane thermal conductivity of Bi$_2$Se$_3$, the experimental error bar can be evaluated to $20\%$.}
\label{tab:results}
\end{table}

\begin{figure}
\includegraphics[width=.50\textwidth]{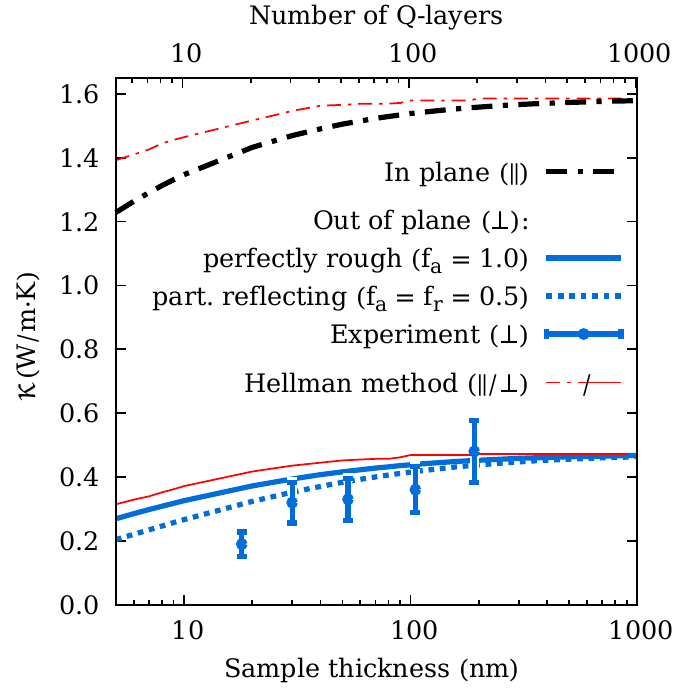}
  \caption{Thermal conductivity in thin films as a function of sample thickness $L$ at 300~K. Experimental data:  lattice thermal conductivity, to be compared with $\kappa$ out-of-plane axis (dashed line). Blue thick line use our theory of Casimir scattering, while red lines use the method of cutting-off phonons of Ref.~\onlinecite{hellman-bite}.}
  \label{fig:tk-thin-noso}
\end{figure}

Thermoreflectance measurements provide the total out-of-plane ($\kappa_\perp$) thermal conductivity which is the sum of the electronic and lattice contributions. As described in section~\ref{electk}, we estimate the lattice contribution after measuring the in-plane electronic conductivity in thin films and estimating the out-of-plane electronic conductivity from the measured conductance anisotropy of bulk \bise{}.

Transport measurements performed attest that our thin films are $n$ doped and present a metallic behavior with a carrier's concentration bracketed by $1-2\times 10^{19}$~cm$^{-3}$, $\mu\sim 300-400$~cm$^2/$V$\cdot$s and $\rho_{\parallel} \sim 1-1.15 $~m$\Omega\cdot$cm at room temperature.

Thus a coarse evaluation of the electronic contribution to the in-plane conductivity $k_{el,\parallel}$ can be given using the Wiedmann-Franz law $k_{el,\parallel} = \mathrm{LT}/\rho_{\parallel}$ with L the Lorentz number, ranging between 2 and $2.2\times 10^{-8}$~V$^2$K$^{-2}$. Consequently $k_{el,\parallel}$ is bracketed between $0.4 - 0.7$~W/m$\cdot$K.

In table~\ref{tab:results} we report the measured values for the total and lattice thermal conductivity, obtained by subtraction the estimated electronic contribution. In Fig.~\ref{fig:tk-thin-noso}, we compare the measured lattice thermal conductivity with the simulations. The agreement is within the experimental errorbar. Including internal reflection effects (dashed line in the figure) does improve the agreement but is not sufficient to explain completely the discrepancy for the smallest slab. This may indicate that a simple Casimir model is not sufficient for such a thin sample, a more detailed description of the interaction of phonon with the surface, including {\bf q} and $\omega$ dependence could improve the agreement. Finally, the laser penetration depth in the sample (around 10~nm) could play a role for the thinnest slabs, although there is no simple way to include it in the simulation.

We have also tested the approach of Ref.~\onlinecite{hellman-bite} (red lines of Fig.~\ref{fig:tk-thin-noso}), which consists in cutting off completely the contribution of phonons that have mean free path $\tau v$ larger than the sample dimension. The behaviour is relatively similar, but the predicted value of $\kappa$ is considerably larger for the smaller samples.

\section{Conclusions}

We calculate the phonon-component in-plane $(\kappa_{\parallel})$ and out-of-plane $(\kappa_{\perp})$ thermal conductivity of Bi$_2$Se$_3$ bulk and films with different thicknesses by using the Boltzmann equation with phonon scattering times obtained from anharmonic third order density functional perturbation theory.  Our results agree with existing measurements on bulk samples\cite{navratil,navratil2} and with our room-temperature thermoreflectance measurements of $\kappa_{\perp}$ on films of thickness ranging from $18$ nm to $191$ nm. 

The calculated thermal conductivity of bulk Bi$_2$Se$_3$ is in excellent agreement with
the experimental data of Ref.\onlinecite{navratil} at all temperatures. While the high temperature limit (e. g. room temperature) is essentially determined by the intrinsic thermal conductivity, the low temperature regime, in the past attributed to Se vacancy, can be very well accounted for by assuming $\approx 100$ ppm Bismuth insertion and $\approx 100$ ppm of Se vacancies. In contrast, Se vacancies alone do not explain the low-T behaviour of the conductivity.

In thin films, we find that the thermal conductivity measured at room temperature, monotonically decreases with reducing film thickness $L$. We can attribute this reduction to incoherent scattering of out-of-plane momentum phonons with the film upper and lower surfaces. Our work outlines the crucial role of sample thinning in reducing the out-of-plane thermal conductivity.

\section{Acknowledgements}
MC acknowledges support from the European Graphene Flagship Core 2 grant number 785219. Computer facilities were provided by CINES, CCRT and IDRIS (projects 907320 and 901202). MM and DF thank Haneen Sameer Abushammala and Heba Hamdan for assistance at the very beginning of thermoreflectance experiments. 

\bibliography{bibliography}

\end{document}